\documentstyle[11pt]{article}

\begin{document}
\centerline {\bf ANYONS BEYOND THE PLANE}
\vskip 1cm
\centerline{
{\bf Edouard Gorbar}$^*$,
{\bf Stefan Mashkevich}$^{*,\dagger}$ \footnote{\it email: mash@phys.unit.no}
{\bf and Sergei Sharapov}$^*$  }
\vskip 1cm
\centerline{$^*$ Institute for Theoretical Physics, Kiev 252143, Ukraine}
\centerline{$^\dagger$ Division de Physique Th\'eorique
\footnote{\it Unit\a'e de Recherche des
Universit\a'es Paris 11 et Paris 6 associ\a'ee au CNRS},  IPN,
  Orsay Fr-91406}
\vskip 1cm
{\bf Abstract:}
We consider particles in three-dimensional space, which have a
certain probability to find themselves in a thin layer (``plane''),
where they are assumed to be well described by a planar Hamiltonian
and are subject to Aharonov-Bohm-type interaction. We demonstrate
that their planar motion is then anyonlike, with the ``effective
statistical parameter'' proportional essentially to the square
of the probability. We also show that charge-flux
composites with solenoids of finite length, provided they are
themselves fermions, can form a bound state in which they behave
like anyons, without any external potential confining them to
a plane.
\vskip 1cm

IPNO/TH 95-20

cond-mat/9503160

March 1995

\vfill\eject

\bigskip
{\bf 1.} As is well-known, fractional statistics is possible
only in two-dimensional space \cite{Leinaas77}
(in one-dimensional space as well, but this is not
our concern here);
this is dictated by general topological arguments.
On the other hand, the real world is three-dimensional,
and two-dimensionality may only be regarded as an
approximation. Therefore for fundamental
particles, fractional
statistics is impossible. On the other hand, it is
intuitively clear that if a particle is confined in
one direction strongly enough, a two-dimensional model
should describe its behavior well enough. Thus, a
question arises about a quantitative description of the
transition from a ``purely three-dimensional'' regime,
with no anyons possible, to a ``quasi-two-dimensional''
one, where they may be considered.

Since no particle itself can be an anyon, in reality it
is only possible to obtain fractional statistics
effectively by means of some interaction.
The most relevant example seems to be the model \cite{Wilczek}
where the ``particles'' (more generally, quasiparticle
excitations) are made of
charges and infinitely thin solenoids and therefore
interact \`a la Aharonov-Bohm.
If the solenoid has an infinite
length, then nothing depends on the third coordinate ($z$) and
the two-dimensional description is exact. If it is much longer
than the domain of motion along the $z$ axis, then this
description may be regarded as a good approximation.
We are going to clarify what happens if one goes beyond
that approximation. We will consider the two-particle
problem, including the calculation of the second virial
coefficient, and the problem of the $N$-particle ground
state. In the second part we will demonstrate that under certain
conditions, anyonic interaction itself may confine
the particles to a plane, forming a bound state.

Anyons are particles
that live on a plane and whose two-particle Hamiltonian is
\begin{equation}
H(R,\Phi;r,\varphi) = H_{\rm CM}(R,\Phi) + H_0(r,\varphi)
+\Delta H (r,\varphi;\alpha),
\end{equation}
where $H_{\rm CM}(R,\Phi)$ is the (irrelevant)
center-of-mass Hamiltonian and
\begin{eqnarray}
H_0(r,\varphi) &\!\!\! = \!\!\!& -\frac{1}{m}\frac{\partial^2}
{\partial r^2} - \frac{1}{mr}\frac{\partial}{\partial r}
-\frac{1}{mr^2}\frac{\partial^2}{\partial \varphi^2}
+\frac{m \omega^2}{4}r^2, \label{2} \\
\Delta H(r,\varphi) &\!\!\! = \!\!\!& \frac{2i\alpha}{mr^2}
\frac{\partial}{\partial\varphi} + \frac{\alpha^2}{mr^2}. \label{3}
\end{eqnarray}
Here the so-called regular gauge \cite{Sen90} has been chosen,
in which the wave function is symmetric (or antisymmetric)
with respect to interchange: $\Psi(r,-\varphi) = \Psi(r,\varphi)$
[$-\Psi(r,\varphi)$, respectively]; $\alpha$ is the statistical
parameter, and a harmonic attraction has been added by hand in order
to discretize the spectrum. The nature of the interaction (\ref{3}),
for example in the context of the fractional quantum Hall effect,
may be understood as follows: The particles are confined to a
plane (the interface between two semiconductors etc.) by an external
potential and being on this plane, they effectively have a magnetic
flux ``glued'' to them; it is then the charge-flux interaction that
is described by $\Delta H(r,\phi)$ (it changes the kinetic angular
momentum by $\alpha$). Imagine now a situation in which the
``plane'' has a finite thickness\footnote{It is to remind of this
thickness that we use the quotation marks.} $2l$ while the particles
can move in the $z$ direction within a range of width $2D$; when a
particle is within the ``plane'', it acquires the flux (stretched
in the $z$ direction) and therefore interacts with the other
particles in the above-described manner, otherwise it does not
interact with them. Clearly, the relevant two-particle Hamiltonian
takes the form
\begin{equation}
H = H^{\perp}(z_1) + H^{\perp}(z_2)
+ H_{CM}(R,\Phi) + H_0(r,\varphi)
+ \theta(l-|z_1|) \theta(l-|z_2|) \Delta H(r,\varphi;\alpha),
\label{4}
\end{equation}
where
\begin{equation}
H^{\perp}(z) = -\frac{1}{2m} \frac{\partial^2}{\partial z^2}
+ V(z),
\label{5}
\end{equation}
$V(z)$ is some confining potential, which determines $D$,
and $\theta(z)$ is the step function.

This model can be said to describe an effective change
of statistics in three dimensions. Indeed,
if $D < l$ (to be more exact, if $V(z) = \infty$ for
$|z|>l$), we arrive to the above-discussed situation of
infinite solenoids, and the planar motion decouples from
the $z$ motion and is purely anyonic. On the contrary,
if $D \gg l$, the last term in (\ref{4}) has essentially
no effect, and one has pure bosons (or fermions). In the
most interesting intermediate case
$D \; \raisebox{-1ex}{$ \stackrel{\textstyle >}{\sim}$} \; l$,
the problem
does not allow for a separation of variables and therefore
in general is not solvable. However, if $|\alpha| \ll 1$,
then it is possible to consider the last term as a
perturbation. The unperturbed solutions are (here and
after we completely ignore the planar center-of-mass
excitations)
\begin{equation}
\psi^{(0)}_{k_1k_2Ln} = \psi^{\perp}_{k_1}(z_1)
\psi^{\perp}_{k_2}(z_2) \psi_{Ln}(r,\varphi),
\label{6}
\end{equation}
where $\psi^{\perp}_k(z)$ are the eigenfunctions of
$H^{\perp}$ and $\psi_{Ln}(r,\varphi)$ are the eigenfunctions of
$H_0$; in particular, for $H_0$ as in (\ref{2}) we have
\begin{equation}
\psi_{Ln}(r,\varphi) = c_{Ln} r^{|L|} e^{iL\varphi}
{}_1\!F_1 \left(-n,|L|+1; -\frac{m\omega}{2}r^2\right) \exp
\left( -\frac{m\omega}{4}r^2 \right),
\label{7}
\end{equation}
and correspondingly
\begin{equation}
E^{(0)}_{k_1k_2Ln} = E^{\perp}_{k_1} + E^{\perp}_{k_2} + (2n+|L|+1)\omega
\label{8}
\end{equation}
($L$ is even/odd for bosons/fermions;
in what follows we will consider bosons, until otherwise stated).
Now, for the two-anyon Hamiltonian, first-order perturbation
theory gives \cite{McCabe,rem} the result which is in fact
correct for any $\alpha$ \cite{Leinaas77,Wilczek},
\begin{equation}
\left\langle \psi_{Ln}(r,\varphi) \right| \Delta H (r,\varphi;\alpha)
\left| \psi_{Ln}(r,\varphi) \right\rangle =
\left\{ \begin{array}{ll} \alpha\omega, & \quad L > 0, \\
|\alpha|\omega, & \quad L=0, \\ -\alpha\omega, & \quad L < 0.
\end{array} \right. \label{9}
\end{equation}
Hence the first-order energy in our problem is easily found to be
\begin{equation}
E^{(1)}_{k_1k_2Ln} = E^{\perp}_{k_1} + E^{\perp}_{k_2}
+ (2n + |L + w_{k_1}w_{k_2}\alpha| + 1)\omega,
\label{10}
\end{equation}
where
\begin{equation}
w_k = \int\limits_{-l}^l \left| \psi^{\perp}_k(z) \right|^2 \, dz
\label{11}
\end{equation}
is the probability that a particle with the wave function
$\psi^{\perp}_k(z)$ stays within the ``plane''.

This result is natural enough, illustrating that the particles
are anyons ``inasmuch as they are in the plane'': The planar
part of the energy corresponds to that of two anyons with
an ``effective statistical parameter'' equal to $\alpha$
times the probability that both particles are in the ``plane''.
Passing from $D \ll l$ to $D \gg l$ will correspond to
passing from $w_{k_{1,2}} \simeq 1$ to $w_{k_{1,2}} \simeq 0$.
Note by the way that for perturbation theory to be
applicable, it is in fact not necessary that $|\alpha| \ll 1$,
it is sufficient that $w_{k_1}w_{k_2}|\alpha| \ll 1$
[the result (\ref{9}) is exact anyway].

As an illustration, let us calculate explicitly the ``planar''
second virial coefficient in this situation, defined as usually
\cite{Arovas},
\begin{equation}
b_2 = \lim_{\omega \to 0} \left( \frac{\lambda T}{\omega}
\right)^2  \left( 1 - 2\frac{{\cal Z}_2}{{\cal Z}_1^2} \right),
\label{v1}
\end{equation}
$\lambda = \sqrt{2\pi/mT}$ being the thermal wavelength.
(Here and further, $\cal Z$ stands for three-dimensional
partition functions, $Z$ and $Z^\perp$ for planar and
perpendicular ones, respectively.) Let the perpendicular potential
be an infinitely deep well:
\begin{equation}
V(z) = \left\{ \begin{array}{ll} 0, & \quad |z| < D, \\
\infty, & \quad |z| > D . \end{array} \right.
\label{v2}
\end{equation}
Then
\begin{equation}
E^{\perp}_k = \frac{\pi^2}{8mD^2}k^2, \qquad
\psi^{\perp}_k(z) = \frac{1}{\sqrt{D}} \sin \frac{\pi k}{2}
\left( 1 + \frac{x}{D} \right),
\label{v3}
\end{equation}
\begin{equation}
w_k = \gamma - \frac{\sin[(1+\gamma)\pi k]}{\pi k},
\label{v4}
\end{equation}
where
\begin{equation}
\gamma = \frac{l}{D}.
\label{v5}
\end{equation}
One has
\begin{eqnarray}
{\cal Z}_1 = Z^{\perp}_1 Z_1
& \!\!\! = \!\!\!& \sum_{k=1}^{\infty}
\exp \left[ - \frac{\pi^2}{8mD^2T} k^2 \right] \times
\sum_{n=0}^{\infty} \sum_{L=-\infty}^{\infty} \exp
\left[ - \frac{(2n+|L|+1)\omega}{T} \right] \nonumber \\
& \!\!\! = \!\!\!& \frac{1}{2} \left[ \theta_3(0,q) - 1 \right] \times
\frac{1}{4 \sinh^2 \frac{\omega}{T}}, \label{v6}
\end{eqnarray}
where
\begin{equation}
q = \exp \left( -\frac{\pi^2}{8mD^2T} \right)
\label{v6a}
\end{equation}
and $\theta_i(z,q), \; i=1,\ldots,4$ are elliptic theta functions.
On the other hand,
\begin{equation}
{\cal Z}_2 = \sum_{\stackrel{\scriptstyle k_1k_2nL}{L \;\rm even}}
\exp \left\{ -\frac{1}{T} \left[ \frac{\pi^2}{8mD^2}(k_1^2 + k_2^2)
+ (2n+|L|+1)\omega + w_{k_1}w_{k_2}\alpha\omega\,\mbox{sign}\:L
\right]\right\} Z_1
\label{v7}
\end{equation}
(for $L=0$, the last term in the square brackets has to be replaced by
$w_{k_1}w_{k_2}|\alpha|\omega$), and expanding to the first
order in $\alpha$, one eventually gets
\begin{equation}
b_2 = \left( -\frac{1}{4} + \mu^2|\alpha| \right) \lambda^2,
\label{v8}
\end{equation}
where
\begin{eqnarray}
\mu &\!\!\! = \!\!\!& \frac{1}{Z^{\perp}_1} \sum_{k=1}^{\infty}
w_k \exp \left( -E_k^{\perp}/T \right) \nonumber \\
& \!\!\! = \!\!\! & \gamma \left[ 1 +
\frac{1- \int_0^1 \theta_4\left( \pi\gamma x/2, q \right) \, dx}
     {\theta_3(0,q) - 1} \right]. \label{v9}
\end{eqnarray}
That is, $\mu$ is the thermal average of the quantity $w_k$, and
again the ``effective statistical parameter'' is proportional to
$\mu^2$, reflecting the fact that two-particle interaction is
involved. Clearly, for $q=0$ (which corresponds to $T=0$)
one has $\mu = w_1 = \gamma + \sin(\pi\gamma)/\pi$, while
for $q=1$ $\;(T=\infty)$ the probability distribution along $z$
becomes uniform and $\mu = \gamma$. Fig.~1 shows
the dependences $\mu(\gamma)$ for these two
limit cases (dashed lines) as well as for an intermediate
case $q=0.75$ (solid line).

Let us now extend the above considerations to the $N$-particle
case. The complete $N$-anyon spectrum is not known, for $N \ge 3$,
but the ground state for small enough $\alpha$ is known: namely, if
\begin{equation}
H_1({\bf r}) = -\frac{{\nabla}^2}{2m} + \frac{m \omega^2}{2}r^2
\label{12}
\end{equation}
is the one-particle (planar) Hamiltonian, $H_n({\bf r}_1, \ldots,
{\bf r}_n; \alpha)$ is the $n$-anyon Hamiltonian in the regular
gauge,  and
\begin{equation}
\Delta H_n({\bf r}_1, \ldots, {\bf r}_n; \alpha) =
H_n({\bf r}_1, \ldots, {\bf r}_n; \alpha) - \sum_{j=1}^{n}
H_1({\bf r}_j),
\label{13}
\end{equation}
then the first-order correction from $\Delta H_n$ to the ground
state of $\sum_j H_1({\bf r}_j),$ with the wave function
$\psi_0({\bf r}_1, \ldots, {\bf r}_n) = c_n \exp
\left( - \frac{m \omega^2}{2} \sum_{j=1}^n {\bf r}^2_j \right)$,
is (see \cite{rem} again)
\begin{equation}
\left\langle \psi_0({\bf r}_1, \ldots, {\bf r}_n) \right|
\Delta H_n ({\bf r}_1, \ldots, {\bf r}_n; \alpha) \left|
\psi_0({\bf r}_1, \ldots, {\bf r}_n) \right\rangle
=\frac{n(n-1)}{2} |\alpha| \omega.
\label{14}
\end{equation}
What will be the generalization of Eq.~(\ref{4}) for this
case? Introduce a quantity
\begin{equation}
\xi^N_{k_1 \ldots k_n} = \theta(l-|z_{k_1}|) \cdots
\theta(l-|z_{k_n}|) \theta(|z_{k_{n+1}}|-l) \cdots
\theta(|z_{k_N}|-l),
\label{15}
\end{equation}
where $k_1, \ldots, k_n$ are any $n$ different numbers from the
set $\{1,\ldots,N\}$ and $k_{n+1},\ldots,k_N$ are the remaining
ones from this set. Thus, $\xi^N_{k_1 \ldots k_n}$ is equal to
1 if and only if the particles with numbers $k_1, \ldots, k_n$
are in the ``plane'' and all the others are out, otherwise it
is equal to 0. Following our assumption that the particles
interact only when they are in the ``plane'', we get
\begin{equation}
H_N = \sum_{j=1}^N H^\perp (z_j) + \sum_{j=1}^N H_1({\bf r}_j)
+ \sum_{n=0}^N \sum_{\{k_1 \ldots k_n\}} \xi^N_{k_1 \ldots k_n}
\Delta H_n ({\bf r}_{k_1}, \ldots, {\bf r}_{k_n}; \alpha),
\label{16}
\end{equation}
where the last sum is performed over all possible choices of $n$
numbers $k_1, \ldots, k_n$, for a given $n$; the number  of terms
in this sum is the binomial coefficient $C^n_N$. Assuming again
that it is possible to consider the last term as a perturbation,
the first-order ground state energy of (\ref{16}) will be
\begin{eqnarray}
E^{(1)}_0 &\!\!\! = \!\!\!& NE^{\perp}_0 + N\omega + \sum_{n=0}^N
C^n_N w_0^n (1-w_0)^{N-n} \frac{n(n-1)}{2} |\alpha|\omega \nonumber \\
&\!\!\! = \!\!\!& NE^{\perp}_0 + N\omega + \frac{N(N-1)}{2}
w_0^2 |\alpha|\omega,
\label{17}
\end{eqnarray}
where again $w_0$ is the probability for a particle to be in the
``plane'',  as in (\ref{11}).
Once again, this corresponds, with the subtraction of the $z$ motion,
to the ground state energy of $N$ anyons with an ``effective
statistical parameter'' $w_0^2\alpha$; this is again natural,
taking into account that upon introducing the ansatz necessary
to make perturbation theory work \cite{rem}, the interaction
Hamiltonian becomes a sum of pair terms. Alternatively, if
$N \gg 1$, one may represent the last term as $\frac{(w_0N)^2}{2}
|\alpha|\omega$, interpreting it as the ground state energy of
$w_0N$ anyons, $w_0N$ being the average quantity of them in the
``plane''.

\bigskip

{\bf 2.} So long we assumed the particles to be confined to a
fixed ``plane'' by an external potential, as it happens, for
example, in the quantum Hall effect; then the whole spectrum
is certainly discrete. Now we are going to show that a two-particle
bound state can as well emerge even when the confining potential
is absent, so that the $z$ motion is free; in this situation
there is no fixed ``plane'', and we suppose that the interaction
is present for $|z_1-z_2|<l$. This would correspond to the particles
being charge-flux composites with solenoid length $l$ and the
function $\Phi(r,z)$, the flux felt by one charge from the other
solenoid when their centers are separated by the vector $({\bf r},z)$,
being approximated by $\phi\cdot\theta(|l|-z)$; dropping the $r$
dependence is justified at least for $r \ll l$, when the flux
lines close far away from the charges. Our idea is simple enough:
If the particles themselves are {\it fermions\/}, then the energy
of the two-particle planar ground state is $2\omega$ outside the
``plane'' but $(2-|\alpha|)\omega$ inside (note that here and
further, $\alpha$ will mean what $(1-\alpha)$ usually means, i.e.~the
deviation from Fermi statistics); now, in three dimensions a state
with energy $E$ such that $(2-|\alpha|)\omega < E < 2\omega$ can
exist---the $z$ dependence of its wave function will be a superposition
of plane waves inside the ``plane'', but exponential damping outside,
due to the right inequality. (See, e.g., \cite{Schult} for a discussion
of an analogous situation, when a bound state exists in a classically
unbound system.)

For an explicit computation, recall the general case: Let
a system have a boundary $z=l$ and thus be characterized
by a Hamiltonian of the form
\begin{equation}
H({\bf r},z) = \left\{ \begin{array}{ll}
H^-({\bf r}) + H^-(z), & \qquad z < l, \\
H^+({\bf r}) + H^+(z), & \qquad z > l. \end{array} \right.
\label{c1}
\end{equation}
Suppose that the following information is available: (i) the
solution of the spectral problem for $H^-({\bf r})$ and
$H^+({\bf r})$, that is the energies and wave functions in
\begin{equation}
H^\pm({\bf r}) \phi^\pm_m ({\bf r})
= {\cal E}^\pm_m \phi^\pm_m ({\bf r}),
\label{c2}
\end{equation}
and (ii) the wave functions $\psi^\pm({\cal E},z)$, {\it for every\/}
$\cal E$, such that
\begin{equation}
H^\pm(z) \psi^\pm ({\cal E},z) = {\cal E} \psi^\pm ({\cal E},z)
\label{c3}
\end{equation}
and that $\int_0^{\pm \infty} |\psi^\pm ({\cal E},z)|^2 \, dz$ is finite.
[If $\cal E$ takes a value within the discrete set of eigenvalues
of the extension
of $H^+(z)$ to $z \in (-\infty,\infty)$, then the integral
$\int_{-\infty}^{\infty} |\psi^+({\cal E},z)|^2 \, dz$ is finite as well,
for any other $\cal E$ it is of course divergent,
the same for $\psi^-({\cal E},z)$.]
Then it is possible to solve the problem (\ref{c1}).
Indeed, since in each of the two regions the variables separate,
the wave function has to be searched for in the form
\begin{equation}
\Psi({\bf r},z) = \left\{ \begin{array}{ll}
\sum\limits_n c^-_n \phi^-_n({\bf r})
\psi^-(E-{\cal E}^-_n,z), & \qquad z < l, \\
\sum\limits_m c^+_m \phi^+_m({\bf r})
\psi^+(E-{\cal E}^+_m,z), & \qquad z > l;
\end{array} \right.
\label{c4}
\end{equation}
by construction, it satisfies the equation
$H({\bf r},z) \Psi({\bf r},z) = E\Psi({\bf r},z)$
and is normalizable. The two boundary conditions take the form
\begin{equation}
\sum_n c^-_n \phi^-_n({\bf r}) \psi^-(E-{\cal E}^-_n,l) =
\sum_m c^+_m \phi^+_m({\bf r}) \psi^+(E-{\cal E}^+_m,l)
\label{c5}
\end{equation}
and the same with $\psi^\pm$ replaced by $\psi^\pm_z
\equiv \partial\psi^\pm/\partial z$.
Scalarly multiplying both these conditions by $\varphi^-_k({\bf r})$,
we get
\begin{equation}
\left\{ \begin{array}{l}
c^-_k \psi^-(E-{\cal E}^-_k,l) = \sum\limits_m c^+_m \Phi_{km}
\psi^+(E-{\cal E}^+_m,l), \\
c^-_k \psi^-_z(E-{\cal E}^-_k,l) = \sum\limits_m c^+_m \Phi_{km}
\psi^+_z(E-{\cal E}^+_m,l), \end{array} \right.
\label{c6}
\end{equation}
where
\begin{equation}
\Phi_{km} = \int \overline{\phi^-_k}({\bf r}) \phi^+_m({\bf r})
\, d^2{\bf r}
\label{c7}
\end{equation}
is the overlap integral, and finally, excluding $c^-_k$,
\begin{equation}
\sum_m A_{km} c^+_m = 0,
\label{c8}
\end{equation}
where
\begin{equation}
A_{km} = \left[
\psi^+(E-{\cal E}^+_m,l) \psi^-_z(E-{\cal E}^-_k,l) -
\psi^+_z(E-{\cal E}^+_m,l) \psi^-(E-{\cal E}^-_k,l) \right] \Phi_{km}.
\label{c9}
\end{equation}
The equation $\; \det A_{kn}=0 \;$ then determines the eigenvalues of $E$.

Let us calculate the ground state in the problem at hand. Symmetry
with respect to $z=0$ is present, hence the ground state wave
function has to be even with respect to $z$ (nodeless), which means
it is enough to consider $z \in [0,\infty)$; we have
\begin{eqnarray}
H^-(z) &\!\!\!=\!\!\!& H^+(z) = 0, \label{g10} \\
H^+({\bf r}) &\!\!\!=\!\!\!& H_0(r,\varphi), \label{g11} \\
H^-({\bf r}) &\!\!\!=\!\!\!& H_0(r,\varphi)
+ \Delta H (r,\varphi;\alpha); \label{g12}
\end{eqnarray}
in view of the aforesaid, one has to choose
\begin{eqnarray}
\psi^+({\cal E},z) &\!\!\!=\!\!\!& \exp (-\sqrt{-2m\cal E}z)
\qquad \qquad ({\cal E} < 0), \label{g13} \\
\psi^-({\cal E},z) &\!\!\!=\!\!\!& \cos (\sqrt{2m\cal E}z)
\qquad \qquad \quad \;\;\; ({\cal E} > 0). \label{g14}
\end{eqnarray}
Again,  let us restrict ourselves to the simplest case $|\alpha| \ll 1$,
whence $\Phi_{km} \simeq \delta_{km} + \alpha X_{km}$; since all terms
in $\; \det A_{km} \,$, except $\prod_k A_{kk}$, will then have at
least a factor $\alpha^2$, we can neglect them. We are to search for
$E$ in the form
\begin{equation}
E = (2-\kappa|\alpha|)\omega,
\label{g15}
\end{equation}
with $0<\kappa<1$, the only possibility is then to have $A_{00}=0$;
now ${\cal E}^-_0 = (2-|\alpha|)\omega, \; {\cal E}^+_0 = 2\omega$,
and an equation for $\kappa$ follows
\begin{equation}
\tan \sqrt{2|\alpha|\xi(1-\kappa)} = \sqrt{\frac{\kappa}{1-\kappa}},
\label{g16}
\end{equation}
where
\begin{equation}
\xi = m\omega l^2.
\label{g17}
\end{equation}
The asymptotic expressions are $\kappa \simeq 2|\alpha| \xi$ for
$|\alpha|\xi\ll1$ and $\kappa\to1$ for $\xi\to\infty$;
the whole dependence $\kappa(\xi)$, for $|\alpha|=0.5$, is
sampled on Fig.~2.
Thus, in this situation it is the ratio of the solenoid length
($l$) and the planar scale ($1/\sqrt{m\omega}$) that matters:
The first one being much greater brings us again to the case of
ideal anyons $(E \to {\cal E}^-_0)$, in the second one, clearly,
the length of the ``tail'' of the wave  function (along $z$) is
much greater than $l$, and $E \to {\cal E}^+_0$, which means
that the particles become essentially free;
however, the bound state exists always,
despite the absence of a confining potential. (Note that the
above-discussed replacement of $\Phi(r,z)$ by the step function
would in fact be good in the first regime only, i.e.~for $\xi\gg 1$;
however, the qualitative picture will remain the same as
described in any case.)

We come to the conclusions that:

1. If the particles (generally speaking, not pointlike) have
a finite probability to find themselves within a ``plane''---a
region where they are subject to anyonic interaction, they may
be regarded, in what concerns their planar motion, as anyons
with the statistical parameter proportional typically to the
square of this probability. This is a natural way of realizing
how one can (effectively) get fractional statistics within
an underlying three-dimensional model;

2. For charge-flux composites with solenoids of finite length
which are fermoions by themselves, the anyonic interaction,
having a character of attraction, can lead to formation of
a bound state even if their longitudinal motion is by no means
restricted.

We wonder whether a specific microscopic model could be found
in which this description would work properly. In particular,
it would be an interesting problem to study in more detail the
influence of possible ``jumps'' of particles (excitations) off
the plane, for example, on the fractional quantum Hall effect,
in particular to observe how the effect disappears when confinement
to the plane becomes less strong.

\bigskip

Acknowledgements: S.M. gratefully acknowledges the hospitality
of the theory division of the IPN at Orsay, where part of this
work was done.

\newpage

{\bf FIGURE CAPTIONS}

Fig.~1. The probability $\mu(\gamma)$ that a particle stays within
the ``plane'', for three cases:
$q=0$ (dashed curve); $q=0.75$ (solid curve); $q=1$ (dashed straight line).

Fig.~2. The function $\kappa(\xi)$ which determines the energy of the
bound state, for $|\alpha|=0.5$.

\begin{thebibliography}{9}
\bibitem{rem} In the $L=0$ case, it is known that perturbation
theory is not directly applicable, since it gives a senseless
infinite energy shift; however, with a certain modification of
the theory itself \cite{McCabe,Comt94},
the result (\ref{9}) will be produced, as well as
(\ref{14}) for the $n$-body case.
\end{thebibliography}
\end{document}